\documentclass[%
 reprint,
groupedaddress,
 amsmath,amssymb,
 aps,twocolumn,
pra,floatfix
]{revtex4-1}

\usepackage{graphicx}
\usepackage{dcolumn}
\usepackage{bm}
\usepackage{dsfont}
\usepackage{blindtext}
\usepackage{mathtools}
\usepackage{lipsum}
\usepackage{xcolor}

\newcommand\varpm{\mathbin{\vcenter{\hbox{%
  \oalign{\hfil$\scriptstyle+$\hfil\cr
          \noalign{\kern-.3ex}
          $\scriptscriptstyle({-})$\cr}%
}}}}

\begin{document}


\title{Engineering of Hong-Ou-Mandel interference with effective noise}

\author{Olli Siltanen}
\email{olmisi@utu.fi}
\affiliation{Turku Centre for Quantum Physics, Department of Physics and Astronomy, University of Turku, FI-20014 Turun yliopisto, Finland}
\affiliation{Laboratory of Quantum Optics, Department of Physics and Astronomy, University of Turku, FI-20014 Turun yliopisto, Finland}

\author{Tom Kuusela}
\affiliation{Turku Centre for Quantum Physics, Department of Physics and Astronomy, University of Turku, FI-20014 Turun yliopisto, Finland}
\affiliation{Laboratory of Quantum Optics, Department of Physics and Astronomy, University of Turku, FI-20014 Turun yliopisto, Finland}

\author{Jyrki Piilo}
\affiliation{Turku Centre for Quantum Physics, Department of Physics and Astronomy, University of Turku, FI-20014 Turun yliopisto, Finland}
\affiliation{Laboratory of Quantum Optics, Department of Physics and Astronomy, University of Turku, FI-20014 Turun yliopisto, Finland}

\date{\today}

\begin{abstract}

The Hong-Ou-Mandel effect lies at the heart of quantum interferometry, having multiple applications in the field of quantum information processing and no classical counterpart. Despite its popularity, only a few works have considered polarization-frequency interaction within the interferometer. In this paper, we fill this gap. Our system of interest is a general biphoton polarization state that experiences effective dephasing noise by becoming entangled with the same photons' frequency state, as the photons propagate through birefringent media. The photons then meet at a beam splitter, where either coincidence or bunching occurs, after which the polarization-frequency interaction continues on the output paths. Alongside performing extensive theoretical analysis on the coincidence probability and different polarization states, we outline multiple interesting applications that range from constructing Bell states to an alternative delayed choice quantum eraser.

\end{abstract}

\maketitle



\section{Introduction}

Hong-Ou-Mandel (HOM) interference has a pivotal role in many-body quantum interferometry, capturing the bosonic nature of photons~\cite{hom_original,hom_review}. When two identical photons meet at a beam splitter, they exit in the same output mode with certainty. In other words, they bunch. As the photons become more distinguishable, coincidence becomes more probable. The experimental signature of the HOM interference, i.e., the coincidence rate dropping to zero with perfectly indistinguishable photons, is referred to as the ``HOM dip"~\cite{hom_review,delayed_choice,maccone,singlet_beam,time_energy_ent,spat_corr,quantum_info,pol_ent,bsm,beatnote,freq_ent,deloc,metro}. The narrower the dip (or, under special conditions, the \textit{peak}~\cite{delayed_choice,hom_peak}), the better the resolution in estimating, e.g., path lengths. Besides its more common use in high-precision quantum metrology~\cite{metro}, Bell state measurement~\cite{hom_peak}, and teleportation~\cite{tele1}, HOM interference has found various new applications in quantum information tasks. HOM interference was recently used, e.g., in measurement-device-independent quantum key distribution protocols~\cite{qkd1,qkd2}.

Although the HOM interference has been extensively studied, there is a surprising shortcoming connecting many of the works. The focus is often in one degree of freedom only, e.g., polarization or frequency. The combined polarization-frequency state is rarely explored, not to mention their possible interaction, i.e., \textit{effective noise}.

Noise is often considered detrimental to open quantum systems~\cite{oqs}. In particular, pure dephasing induces quantum-to-classical transition by converting the coherences of a given system to quantum correlations between the system and its environment, which together form a closed system~\cite{zurek,suter}. Fortunately, decoherence can be controlled by means of reservoir engineering, i.e., manipulating the environment degrees of freedom and the initial system-environment correlations~\cite{good_for}. In some cases, such systems undergo non-Markovian dynamics and outdo their Markovian counterparts, e.g., in the Deutsch–Jozsa algorithm~\cite{dj_alg} and superdense coding~\cite{superdense}. Furthermore, there are cases in which the mere presence of noise can be utilized, non-Markovian or not; in~\cite{probing}, it was experimentally demonstrated that even with an unknown system-environment (or system-probe) coupling it is possible to obtain nontrivial information about the system of interest by measuring the coupled probe.

Commonly in the linear optical framework, polarization is the open system, frequency (or temporal) modes represent the environment, and their interaction is implemented controllably in a birefringent medium~\cite{superdense,probing,sim1,sim2,sim3,sim4,sim5,M_to_NM,shaham1,shaham2,nonlocal_memory,photonic_real,tele2,tele3,shaham3,synthetic,sina,partitions,mz_nm}. What is unitary evolution for the total system containing both polarization and frequency, becomes nonunitary when focusing only on the open system by tracing over the environment. So, despite the lack of more orthodox environment such as heat bath, the phase information of the polarization vanishes, corresponding to effective dephasing noise. In~\cite{mz_nm}, dephasing was studied in the context of single-photon interference from the point of view of open quantum systems. Here, we consider two photons in a HOM interferometer, analyze the effects of dephasing noise before and after the beam splitter, and emphasize the coincidence/bunching paradigm, going well beyond previous works dealing with HOM interference and birefringence~\cite{bire1,bire2,bire3,bire4,bire5,bire6,bire7}. The generality of our model allows us to explore the whole polarization-frequency space spanned by the biphoton system, the rich polarization dynamics arising from the ``open system interferometer", and the role of initial correlations. It also enables us to optimize initial conditions for various applications that range from constructing Bell states to dynamical delayed choice erasure.

The paper is structured as follows: In Section II, we present a general version of our model, which we will then study in more detail in the following sections. In Sec. III, we calculate the coincidence probability and analyze how different noise configurations affect the HOM dip. From there on, we focus on a longer and more realistic path difference, and how noise applied after the beam splitter can compensate it in different dynamical tasks. In Sec. IV, we describe a robust method to construct pure Bell states with dephasing, after which we concentrate on single-photon polarization dynamics at one of the output modes only. In Sec. V, we show how different initial conditions can be estimated with dephasing combined with a sufficient dead time of the photodetector. Distinguishing between coincidence and bunching events without an actual coincidence counter is considered in Sec. VI. Section VII concludes the paper.

\section{The model}

\begin{figure}[t!]
\centering
\includegraphics[width=\linewidth]{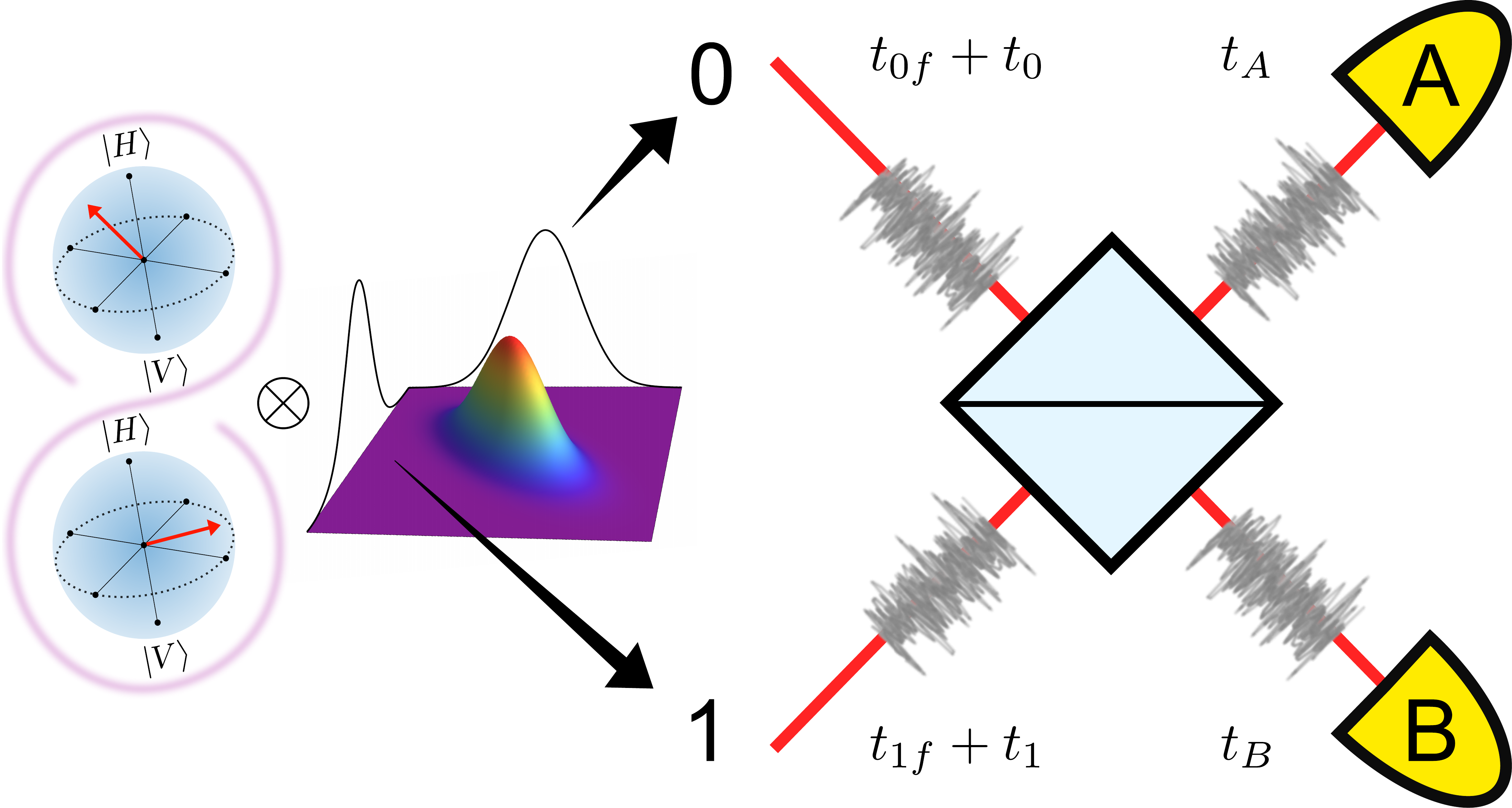}
\caption{A schematic picture of our system of interest (biphoton polarization), its environment (biphoton frequency), and their interaction within a HOM interferometer with dephasing noise both before and after the beam splitter ($t_0$, $t_1$, $t_A$, $t_B$). Free evolution is only considered before the beam splitter ($t_{0f}$, $t_{1f}$).}
\label{Scheme}
\end{figure}

Our system of interest consists of two photons initially on their own paths, which are labeled with 0 and 1. The photons are sent to a beam splitter, from which they then go to Alice (path A), Bob (path B), or one to each. Here, polarization-frequency interaction is considered both before and after the beam splitter, while free evolution is only considered on paths 0 and 1 (see Fig.~\ref{Scheme}). This is because we are interested in the coincidence probability and the polarization dynamics after interference, to which free evolution on paths A and B does not contribute.

For simplicity, we assume that the initial polarization and frequency states are pure and thus uncorrelated. The initial state of the bipartite polarization-frequency system is then
\begin{equation}
\begin{split}
|\psi_{in}\rangle=\Big[&C_{HH}\int d\omega_0d\omega_1g(\omega_0,\omega_1)\hat{a}^\dagger_H(\omega_0)\hat{b}^\dagger_H(\omega_1)\\
+&C_{HV}\int d\omega_0d\omega_1g(\omega_0,\omega_1)\hat{a}^\dagger_H(\omega_0)\hat{b}^\dagger_V(\omega_1)\\
+&C_{VH}\int d\omega_0d\omega_1g(\omega_0,\omega_1)\hat{a}^\dagger_V(\omega_0)\hat{b}^\dagger_H(\omega_1)\\
+&C_{VV}\int d\omega_0d\omega_1g(\omega_0,\omega_1)\hat{a}^\dagger_V(\omega_0)\hat{b}^\dagger_V(\omega_1)\Big]|0_{ab}\rangle,
\end{split}
\label{initial_state}
\end{equation}
where $C_{\lambda\lambda'}$ and $g(\omega_0,\omega_1)$ are the probability amplitudes for the two photons to be in the polarization state $|\lambda\lambda'\rangle$ ($\lambda,\lambda'\in\{H,V\}$) and the (angular) frequency state $|\omega_0\omega_1\rangle$, respectively. $a_\lambda^\dagger(\omega_0)$ is the creation operator corresponding to the polarization component $\lambda$ and the frequency eigenstate $|\omega_0\rangle$ on path 0 while $b_\lambda^\dagger(\omega_1)$ is related to path 1, and $|0_{ab}\rangle$ is the vacuum state. The creation operators transform in free evolution and the polarization-frequency interaction according to
\begin{equation}
\begin{cases}
\hat{a}^\dagger_\lambda(\omega_0)\mapsto e^{i(t_{0f}+n_{0\lambda} t_0)\omega_0}\hat{a}^\dagger_\lambda(\omega_0),\\
\hat{b}^\dagger_\lambda(\omega_1)\mapsto e^{i(t_{1f}+n_{1\lambda} t_1)\omega_1}\hat{b}^\dagger_\lambda(\omega_1),
\end{cases}
\label{free_ev}
\end{equation}
where $t_{0f}$ $(t_{1f})$ is the duration of free evolution on path 0 (1), $t_0$ $(t_1)$ is the total interaction time on path 0 (1), and $n_{0\lambda}$ $(n_{1\lambda})$ is the refractive index of a birefringent medium on path 0 (1) corresponding to the polarization component $\lambda$. The action of the beam splitter reads
\begin{equation}
\begin{cases}
\hat{a}^\dagger_\lambda(\omega_0)\mapsto[\hat{a}^\dagger_\lambda(\omega_0)+\hat{b}^\dagger_\lambda(\omega_0)]/\sqrt{2},\\
\hat{b}^\dagger_\lambda(\omega_1)\mapsto[\hat{a}^\dagger_\lambda(\omega_1)-\hat{b}^\dagger_\lambda(\omega_1)]/\sqrt{2},
\end{cases}
\label{bs}
\end{equation}
where, on the right-hand side, $\hat{a}_\lambda^\dagger(\omega_j)$ refers to path A and $\hat{b}_\lambda^\dagger(\omega_j)$ to path B. Finally, the dephasing implemented by Alice and Bob on their paths after the beam splitter is described by the transformations
\begin{equation}
\begin{cases}
\hat{a}^\dagger_\lambda(\omega_0)\mapsto e^{in_{A\lambda}t_A\omega_0}\hat{a}^\dagger_\lambda(\omega_0),\\
\hat{b}^\dagger_\lambda(\omega_1)\mapsto e^{in_{B\lambda}t_B\omega_1}\hat{b}^\dagger_\lambda(\omega_1).
\end{cases}
\label{dephasing}
\end{equation}
\\

Plugging Eqs.~\eqref{free_ev}--\eqref{dephasing} into Eq. \eqref{initial_state} gives us the output state
\begin{equation}
\begin{split}
|\psi_{out}\rangle=&\sum_{\lambda,\lambda'=H,V}\frac{C_{\lambda\lambda'}}{2}\int d\omega_0d\omega_1g(\omega_0,\omega_1)\\
&\times e^{i[(t_{0f}+n_{0\lambda}t_0)\omega_0+(t_{1f}+n_{1\lambda'}t_1)\omega_1]}\\
&\times[e^{i(n_{A\lambda}t_A\omega_0+n_{A\lambda'}t_A\omega_1)}\hat{a}^\dagger_\lambda(\omega_0)\hat{a}^\dagger_{\lambda'}(\omega_1)\\
&-e^{i(n_{A\lambda}t_A\omega_0+n_{B\lambda'}t_B\omega_1)}\hat{a}^\dagger_\lambda(\omega_0)\hat{b}^\dagger_{\lambda'}(\omega_1)\\
&+e^{i(n_{B\lambda}t_B\omega_0+n_{A\lambda'}t_A\omega_1)}\hat{b}^\dagger_\lambda(\omega_0)\hat{a}^\dagger_{\lambda'}(\omega_1)\\
&-e^{i(n_{B\lambda}t_B\omega_0+n_{B\lambda'}t_B\omega_1)}\hat{b}^\dagger_\lambda(\omega_0)\hat{b}^\dagger_{\lambda'}(\omega_1)]
|0_{ab}\rangle,
\end{split}
\label{output_state}
\end{equation}
where we still haven't considered if both of the photons go to Alice or Bob (bunching), or one to each (coincidence). The projection operator corresponding to Alice receiving only one photon is given by
\begin{equation}
\hat{P}_A=\int d\omega\hat{a}^\dagger_H(\omega)|0_a\rangle\langle0_a|\hat{a}_H(\omega)+\int d\omega\hat{a}^\dagger_V(\omega)|0_a\rangle\langle0_a|\hat{a}_V(\omega),
\label{c_proj}
\end{equation}
while the projector corresponding to Alice receiving both of the photons is
\begin{equation}
\begin{split}
\hat{P}_{AA}=\frac{1}{2}\sum_{\lambda,\lambda'=H,V}&\int d\omega_0d\omega_1\hat{a}^\dagger_\lambda(\omega_0)\hat{a}^\dagger_{\lambda'}(\omega_1)\\
&\times|0_a\rangle\langle0_a|\hat{a}_\lambda(\omega_0)\hat{a}_{\lambda'}(\omega_1)\otimes|0_b\rangle\langle0_b|,
\end{split}
\label{b_proj}
\end{equation}
and similarly for Bob.

\section{Coherent engineering of the coincidence probability}

Here, we calculate the coincidence probability for our system and discuss a few scenarios in which the path difference is zero, i.e., $t_{0f}-t_{1f}=0$, to see how dephasing alone affects the HOM dip and how it differs from the conventional case. Note that throughout this paper, with ``path difference", we refer to the difference in free evolution on paths 0 and 1, although dephasing is applied upon the same paths.

Let us simplify the analysis by assuming that $g(\omega_0,\omega_1)=g(\omega_1,\omega_0)$ and using the bivariate Gaussian
\begin{equation}
|g(\omega_0,\omega_1)|^2=\frac{1}{2\pi\sqrt{\det C}}e^{-\frac{1}{2}(\vec{\omega}-\langle\vec{\omega}\rangle)^TC^{-1}(\vec{\omega}-\langle\vec{\omega}\rangle)}
\label{Gauss}
\end{equation}
with the vectors $\vec{\omega}=(\omega_0,\omega_1)^T$ and $\langle\vec{\omega}\rangle=(\langle\omega_0\rangle,\langle\omega_1\rangle)^T$, and the covariance matrix elements $C_{ij}=\langle\omega_i\omega_j\rangle-\langle\omega_i\rangle\langle\omega_j\rangle$. For further simplicity, we assume that $\langle\omega_0\rangle=\langle\omega_1\rangle=\mu$ and $C_{00}=C_{11}=\sigma^2$, i.e., that the mean frequency $\mu$ and the variance $\sigma^2$ are the same for both photons. The correlation coefficient $K=C_{01}/C_{00}$ quantifies the initial frequency correlations and satisfies $|K|\leq1$. The joint spectrum~\eqref{Gauss} is a good approximation of the one produced in spontaneous parametric down-conversion with the pump frequency $2\mu$~\cite{nonlocal_memory,photonic_real,tele2,tele3,partitions,dfs,humble1,humble2}. Large enough ratio $\mu/\sigma$ guarantees that taking the integrals over frequency from $-\infty$ to $\infty$ is also a good approximation.

The coincidence probability, i.e., the probability for both Alice and Bob receiving one photon, then becomes
\begin{widetext}
\begin{align}
P_c&=\text{tr}[\hat{P}_A\otimes\hat{P}_B|\psi_{out}\rangle\langle\psi_{out}|\hat{P}_A\otimes\hat{P}_B]
\label{P_c_1}\\
&=\langle\psi_{out}|\hat{P}_A\otimes\hat{P}_B|\psi_{out}\rangle
\label{P_c_2}\\
&=\frac{1}{2}\Big\{1-|C_{HH}|^2e^{-(1-K)(\Delta\tau_f+\Delta\tau_{HH})^2}-|C_{VV}|^2e^{-(1-K)(\Delta\tau_f+\Delta\tau_{VV})^2}
\label{P_c_3}\\
&\nonumber\hspace{32.5pt}-2|C_{HV}||C_{VH}|e^{-\frac{1}{2}[(\Delta\tau_f+\Delta\tau_{HH})^2-2K(\Delta\tau_f+\Delta\tau_{HH})(\Delta\tau_f+\Delta\tau_{VV})+(\Delta\tau_f+\Delta\tau_{VV})^2]}\cos[\eta(\tau_0-\tau_1)+\theta_{HV}-\theta_{VH}]\Big\},
\end{align}
\end{widetext}
where $\Delta\tau_f=\sigma(t_{0f}-t_{1f})$, $\Delta\tau_{\lambda\lambda}=\sigma(n_{0\lambda}t_0-n_{1\lambda}t_1)$, $\tau_{j}=\sigma\Delta n_jt_j=\sigma(n_{jH}-n_{jV})t_j$, $j\in\{0,1\}$, $\eta=\mu/\sigma$, and $\theta_{HV}$ ($\theta_{VH}$) is the phase of $C_{HV}$ ($C_{VH}$). The exponential functions that are weighted by the probabilities $|C_{\lambda\lambda}|^2$ depict the temporal differences between the input paths' $\lambda$ components. The cross-terms, on the other hand, experience oscillation with the frequency $\mu$. The $\mu$-dependency of Eq.~\eqref{P_c_3} comes solely from dephasing. In this setting, $P_c$ is not affected by the dephasing on paths A and B. The receiver-specific bunching probability, i.e., the probability for Alice (Bob) to receive both of the photons, is simply $P_b^{A(B)}=(1-P_c)/2.$

We note that, when the two photons entering the beam splitter are separable in polarization and differ only by their path lengths, Eq.~\eqref{P_c_3} reduces to
\begin{equation}
P_c=\frac{1}{2}\Big[1-e^{-(1-K)\Delta\tau_f^2}\Big],
\label{classical_HOM}
\end{equation}
which describes the classical HOM dip of a bivariate Gaussian as a function $\Delta\tau_f$~\cite{hom_original,deloc,beatnote}. The narrowest HOM dip is given by frequency-anticorrelated photons ($K=-1$), while the dip ``fattens" as $K$ approaches 1 and the frequencies become positively correlated~\cite{maccone,pos_corr}. This is due to HOM-type interference being \textit{differential-frequency} interference~\cite{fattening}. That is, the Gaussian~\eqref{Gauss} can be decomposed into the form $|g_+(\omega_0+\omega_1)|^2|g_-(\omega_0-\omega_1)|^2$, where only the latter term contributes to $P_c$. $|g_-(\omega_0-\omega_1)|^2$ approaches the delta function as $K\to1$ but is well-defined at $K=-1$.

On the other hand, with no restrictions on the initial polarization state's separability, but fixing $\Delta\tau_f=0$ (and keeping the noise configurations on paths 0 and 1 still the same), we get
\begin{equation}
P_c=\frac{1}{2}|C_{HV}-C_{VH}|^2.
\label{singlet}
\end{equation}
With the singlet state $|\Psi^-\rangle=(|HV\rangle-|VH\rangle)/\sqrt{2}$, it is possible to obtain $P_c=1$, as was experimentally demonstrated in~\cite{hom_peak}.

As one might expect---and clearly sees from Eq.~\eqref{P_c_3}---dephasing only affects the probabilities when there is some imbalance between the two input modes. The coincidence probabilities~\eqref{classical_HOM} and~\eqref{singlet} can be obtained even in noisy circumstances if the dephasing channels are just identical, which goes to show that the HOM interference also accounts for polarization-frequency correlations. The following remark supports this claim. Should the initial polarization state be $|++\rangle=(|HH\rangle+|HV\rangle+|VH\rangle+|VV\rangle)/2$, full decoherence results in fully mixed polarization state and, with equal path lengths and interaction times, the coincidence probability $P_c=0$ [see Eq.~\eqref{singlet}]. On the other hand, one can construct the same polarization state entering the beam splitter by statistical mixing and obtain $P_c=1/4$. This is due to having no polarization-frequency correlations.

\begin{figure}[t!]
\centering
\includegraphics[width=.9\linewidth]{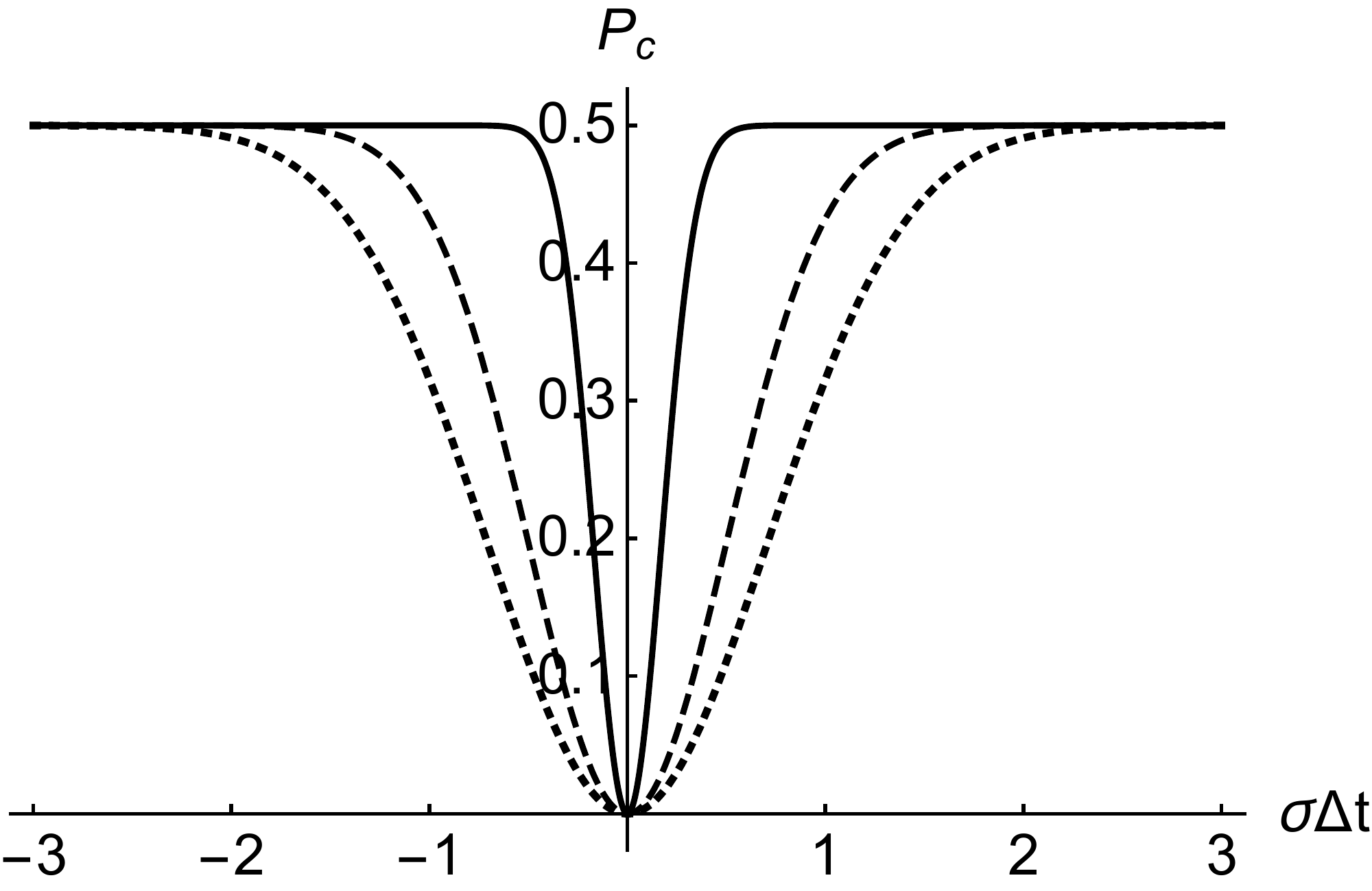}
\caption{The HOM dips related to free evolution with $K=0$ (dotted) and $K=-1$ (dashed), and dephasing with $n_\lambda=2.903$ and $K=-1$ (solid) as functions of the scaled path/interaction difference $\sigma\Delta t$.}
\label{Dips}
\end{figure}

We now consider two scenarios in which dephasing does have some effect on $P_c$ in more detail. In both of them $\Delta\tau_f=0$. First, we have the initial polarization state $|\lambda\lambda\rangle$ and $n_{0\lambda}=n_{1\lambda}=n_\lambda$, so that the coincidence probability becomes
\begin{equation}
P_c=\frac{1}{2}\Big[1-e^{-(1-K)\sigma^2n_\lambda^2(t_0-t_1)^2}\Big].
\label{product_state}
\end{equation}
Because $n_\lambda>1$, we can reach narrower HOM dip in this case than with Eq.~\eqref{classical_HOM}, meaning that the differences in thicknesses of two same birefringent media can be estimated with better resolution than the path differences related to free evolution. This result can be applied, e.g., in high-precision measurement of the birefringent material's thermal expansion or the refractive index's dependency on different factors like temperature and stress. In Fig.~\ref{Dips}, we have plotted the ``traditional" $P_c$ [Eq.~\eqref{classical_HOM}] with $K=0$ and $K=-1$, and the ``noisy" $P_c$ [Eq.~\eqref{product_state}] with $K=-1$ and $n_\lambda=2.903$ (the refractive index of extraordinary ray for rutile at 589.3 nm~\cite{rutile}).

Secondly, we consider the case where the media are equally thick but whose fast axes are perpendicular to each other, i.e., $\sigma(n_{0H}-n_{0V})t_0=\sigma(n_{1V}-n_{1H})t_1=\tau$. With these choices, the coincidence probability becomes
\begin{equation}
\begin{split}
P_c&=\frac{1}{2}\Big[1-\big(|C_{HH}|^2+|C_{VV}|^2\big)e^{-(1-K)\tau^2}\\
&-2|C_{HV}||C_{VH}|e^{-(1+K)\tau^2}\cos(2\eta\tau+\theta_{HV}-\theta_{VH})\Big],
\end{split}
\label{perp_noise}
\end{equation}
which decreases with increasing interaction time $\tau$; $P_c$ approaches 1/2 independently of the initial polarization state when $\tau\to\infty$ and $|K|\neq1$. This again illustrates the significance of the polarization-frequency correlations. Interestingly, $K=1$ protects the polarization subspace $\text{span}(\{|HH\rangle,|VV\rangle\})$ from dephasing, while $K=-1$ protects $\text{span}(\{|HV\rangle,|VH\rangle\})$---excluding rotation dictated by the cosine term---which is the exact opposite of the usual case~\cite{dfs}. Note that the oscillations and $K=-1$ allow one to have $P_c=1$ with any initial polarization state of the form $(|HV\rangle+e^{i\phi}|VH\rangle)/\sqrt{2}$. In general, different orientations of the birefringent media allow for flexible engineering of the HOM dip.

\section{Constructing Bell states with delay-compensating noise}

The condition $\Delta\tau_f=0$ is required, e.g., in constructing Bell states by guiding two photons carrying the polarization qubits $|H\rangle$ and $|V\rangle$ simultaneously into a beam splitter~\cite{bell}. In reality, exact control over the path lengths is extremely demanding. Hence, we shall focus on $\Delta\tau_f\neq0$ (and even $|\Delta\tau_f|>>0$) in the rest of the paper. Furthermore, we will consider dephasing on the exit paths only and, in this section, show how it can be used to construct Bell states. A delay-independent method to prepare Bell states with beam splitters and half-wave plates was demonstrated quite recently in~\cite{bsm}. We mitigate the problem with path difference in an alternative fashion that requires fewer resources and---with little modification---produces the same biphoton polarization state regardless of whether a coincidence or bunching occurs. The idea behind our method is rather simple; horizontally and vertically polarized photons are guided into a beam splitter at different times, after which dephasing noise is implemented at the exit paths. The photons propagate in the dephasing channels with different velocities and overlap at some point in time, becoming fully entangled. With dephasing, the temporal overlap of the photons is considerably easier to achieve than in free air.

In principle, our method would work as well with dephasing applied before the beam splitter. However, controlling the interaction times after the beam splitter may be more convenient for spatially separated parties. In addition, our results show that the photons do not have to overlap at the beam splitter. A special case of our scheme was already demonstrated in~\cite{bire2}, which we go well beyond by considering different noise configurations, free correlation coefficient, and local Bell states corresponding to the bunching events.

Let us fix $C_{HV}=1$ and $t_0=t_1=0$ in Eq.~\eqref{output_state}. The state after the beam splitter and the interaction times $t_A$ and $t_B$ can now be written as
\begin{equation}
\begin{split}
|\psi_{out}\rangle=&\frac{1}{2}\int d\omega_0d\omega_1g(\omega_0,\omega_1)e^{i(t_{0f}\omega_0+t_{1f}\omega_1)}\\
&\times[e^{i(n_{AH}t_A\omega_0+n_{AV}t_A\omega_1)}\hat{a}^\dagger_H(\omega_0)\hat{a}^\dagger_V(\omega_1)\\
&-e^{i(n_{AH}t_A\omega_0+n_{BV}t_B\omega_1)}\hat{a}^\dagger_H(\omega_0)\hat{b}^\dagger_V(\omega_1)\\
&+e^{i(n_{BH}t_B\omega_0+n_{AV}t_A\omega_1)}\hat{b}^\dagger_H(\omega_0)\hat{a}^\dagger_V(\omega_1)\\
&-e^{i(n_{BH}t_B\omega_0+n_{BV}t_B\omega_1)}\hat{b}^\dagger_H(\omega_0)\hat{b}^\dagger_V(\omega_1)]
|0_{ab}\rangle.
\end{split}
\label{HV}
\end{equation}
In the case of coincidence (whose probability is $P_c=1/2$ due to the orthogonal initial states), the polarization state shared by Alice and Bob is (see the detailed derivation in Appendix A)
\begin{equation}
\varrho_c(t_A,t_B)=\frac{1}{2}
\begin{pmatrix}
0 & 0 & 0 & 0 \\
0 & 1 & \Lambda_c(t_A,t_B) & 0 \\
0 & \Lambda_c(t_A,t_B)^* & 1 & 0 \\
0 & 0 & 0 & 0
\end{pmatrix},
\label{coin_Bell}
\end{equation}
where $\Lambda_c(t_A,t_B)=-\exp\Big\{i\eta(\tau_A-\tau_B)-\frac{1}{2}\Big[(\tau_A+\Delta\tau_f)^2-2K(\tau_A+\Delta\tau_f)(\tau_B+\Delta\tau_f)+(\tau_B+\Delta\tau_f)^2\Big]\Big\}$
is the nonlocal decoherence function with $\tau_j=\sigma\Delta n_jt_j=\sigma(n_{jH}-n_{jV})t_j$, $j\in\{A,B\}$. In the case of bunching on Alice's side (whose probability is $P_b^A=1/4$), the polarization state is (see Appendix A for details)
\begin{equation}
\varrho_b^A(t_A)=\frac{1}{2}
\begin{pmatrix}
0 & 0 & 0 & 0 \\
0 & 1 & \Lambda_b^A(t_A) & 0 \\
0 & \Lambda_b^A(t_A)^* & 1 & 0 \\
0 & 0 & 0 & 0
\end{pmatrix},
\label{bunch_Bell}
\end{equation}
where $\Lambda_b^A(t_A)=-\Lambda_c(t_A,t_A)=e^{-(1-K)(\tau_A+\Delta\tau_f)^2}$, and similarly for Bob. From Eqs.~\eqref{coin_Bell} and~\eqref{bunch_Bell}, it is clear that no noise is needed if $\Delta\tau_f=0$. In such situation, Alice and Bob either share the Bell state $|\Psi^-\rangle$ (which can be changed to other Bell states by either Alice or Bob applying proper Pauli operators) or have $|\Psi^+\rangle=(|HV\rangle+|VH\rangle)/\sqrt{2}$ completely on their side. However, as mentioned earlier, calibrating $\Delta\tau_f$ in such precision is very difficult. Due to the birefringences $\Delta n_j$ being close to zero, the time scales of dephasing are more accessible than the time scales of free evolution.

\begin{figure}[t!]
\centering
\includegraphics[width=\linewidth]{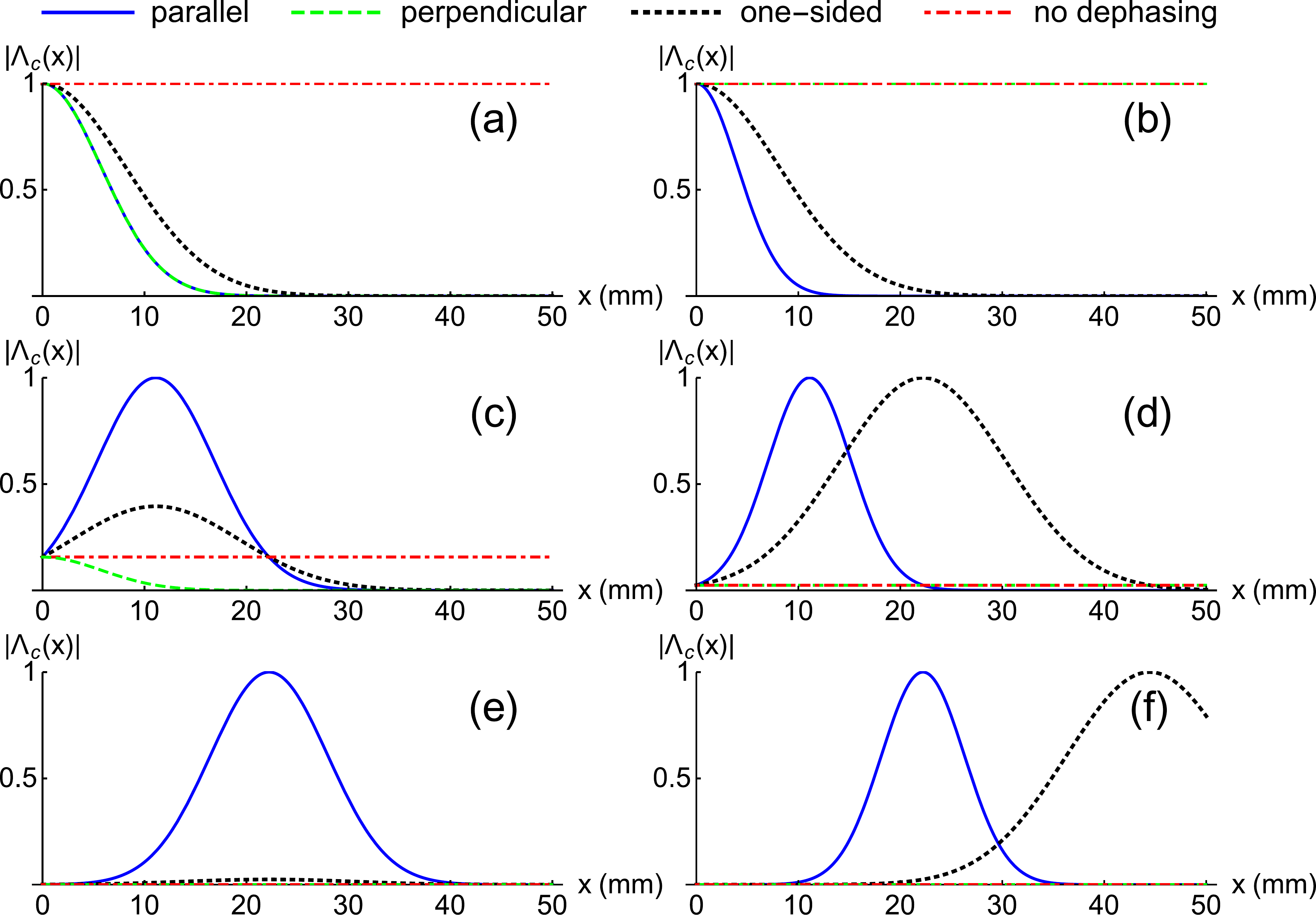}
\caption{(Color online) Absolute values of the decoherence function $\Lambda_c$ as functions of the thickness of quartz, when the fast axes on Alice and Bob's sides are aligned (blue, solid), perpendicular (green, dashed), when only Alice or Bob implements noise (black, dotted), and when there is no noise on either side (red, dash-dotted). The path differences are (a), (b): 0 mm; (c), (d): $-0.1$ mm; (e), (f): $-0.2$ mm. The correlation coefficients $K$ are (a), (c), (e): 0; (b), (d), (f): $-1$. In each panel, we have used the bandwidth of 650 GHz and $\Delta n=0.009$.}
\label{Lambdas}
\end{figure}

There are different protocols that Alice and Bob can agree on. They can apply the same amount of noise ($\tau_A=\tau_B$), ``opposite amounts" of noise ($\tau_A=-\tau_B$, i.e., Bob rotates his birefringent medium by 90 degrees), only one of them applies noise, and so on. In Fig.~\ref{Lambdas}, we have plotted the absolute values of $\Lambda_c(t_A,t_B)$ corresponding to the three protocols described above with different values of $\Delta\tau_f$ and $K$. The blue curves also describe $\Lambda_b^j(t_j)$. For comparison, we have also plotted the constant ``decoherence function" $e^{-(1-K)\Delta\tau_f^2}$, which is related to the conventional case without noise.

From Fig.~\ref{Lambdas} we see that, to reach the maximum purity of $|\Psi^-\rangle$, Alice and Bob applying the same amount of noise requires the least of it, and that the maximum value of $|\Lambda_c(t_j,t_j)|=|\Lambda_b^j(t_j)|$ is independent of $K$. With $K=-1$, it is striking that Alice or Bob alone---with the phase difference of $\pi$---can fully entangle their shared polarization state and, while doing so, purify the state $|\Psi^+\rangle$ related to bunching in the half-way. Perpendicular axes and $K=-1$ result in the decoherence-free subspace $\text{span}(\{|HV\rangle,|VH\rangle\})$ that, in this case, is highly sensitive to $\Delta\tau_f$. Comparing the rows with each other, it becomes clear what we mean by the ``accessibility of time scales". With our parameter choices, it takes less than 0.2 mm for the coherence terms in the usual case to reach zero, while the coherence length related to our method is of the order of 10 mm. 

To prepare the same Bell state independently of coincidence and bunching (up to a global phase factor), Alice and Bob need to align their fast axes and use equal interaction times. With $t_A=t_B=-\Delta t_f/\Delta n$, Alice and Bob either share the singlet state $|\Psi^-\rangle$ or have $|\Psi^+\rangle$ completely on their side. However, if either Alice or Bob operates with $\sigma_z$ after the noise, the state becomes $|\Psi^+\rangle$ in all cases; should a coincidence occur, $\sigma_z$ operates only once and transforms $|\Psi^-\rangle$ into $|\Psi^+\rangle$. Bunching, on the other hand, translates into operating with $\sigma_z$ to either both of the photons or neither one, leaving the state as it was. In this special case, we obtain the success rate of 1.

Clearly, our method can also be applied in estimating different parameters when there is no more information about them in the coincidence rate. Alice and Bob can, e.g., estimate $K$ and $\Delta\tau_f$ by monitoring the nonlocal dephasing dynamics. In the case of parallel media, the point of time corresponding to $|\Lambda_c(t_j,t_j)|=|\Lambda_b^j(t_j)|=1$ is directly proportional to $\Delta\tau_f$, while the width of the decoherence functions is related to $K$. In the following section, we will propose an alternative method for this task that, against intuition, does not require communication between Alice and Bob.

\section{Local evaluation of initial correlations and path difference}

\begin{figure}[t!]
\centering
\includegraphics[width=\linewidth]{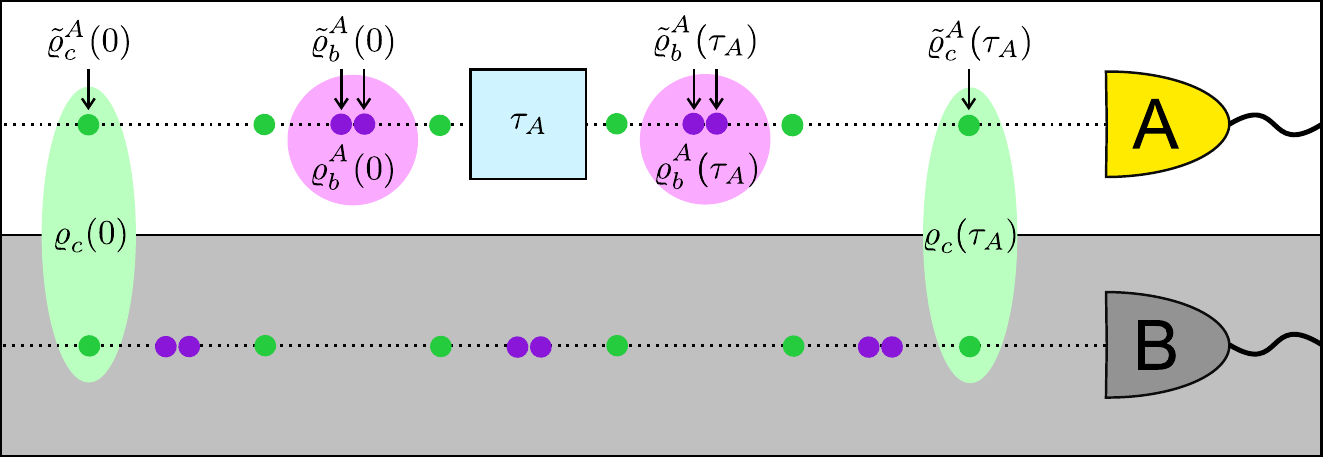}
\caption{Different single and biphoton polarization states after the beam splitter of a HOM interferometer, before and after the polarization-frequency interaction. We stress that this figure is purely schematic, as it is not until the detectors that the coincidence and bunching events actualize. For clarity, the quarter-wave plate, half-wave plate, and polarizer needed in single-photon polarization tomography on Alice's side are not shown in this figure.}
\label{Local_scheme}
\end{figure}

Let us now focus on Alice alone and what she can conclude from single-photon process tomography data without communicating with Bob (the opposite roles are handled similarly). In particular, they do not share a coincidence counter, so that Alice cannot distinguish between the coincidence (\textit{c}) photons and bunched (\textit{b}) photons. The polarization state that Alice can construct with tomographic measurements is a mixed state consisting of \textit{c} and \textit{b} photons (see Fig.~\ref{Local_scheme}). We make the important assumption that $|\Delta\tau_f|$ is great enough to temporally separate the \textit{b} photons so that their contributions on the state tomography can be treated separately. First, we show that, when using an ideal photodetector with close-to-zero dead time, i.e., the time that the detector is off after each detection, $K$ and $\Delta\tau_f$ cannot be detected by monitoring the local single-qubit dynamics. Then, we show that increasing the dead time allows Alice to locally evaluate the correlations and path-difference. It is important to notice here that Alice performs \textit{single}-photon tomography although sometimes, as a result of bunching, she receives a biphoton polarization state. We also note that our scheme cannot distinguish between classical and quantum correlations, while, e.g., Refs.~\cite{discord1,discord2} introduced a local method to detect bipartite quantum discord.

For simplicity---and reasons that we will discuss later---we assume that the initial polarization state is separable and that the local states are identical, so that only $K$ and $\Delta\tau_f$ affect the coincidence probability given by Eq.~\eqref{classical_HOM}. Then, the polarization states carried by the \textit{c} and \textit{b} photons are, respectively (see the detailed derivations in Appendix B),
\begin{equation}
\tilde{\varrho}_c^A(\tau_A)=\begin{pmatrix}
|C_H|^2 & C_HC_V^*\kappa_-(\tau_A) \\
C_H^*C_V\kappa_-(\tau_A)^* & |C_V|^2
\end{pmatrix}
\label{loc_id_rho_c}
\end{equation}
and
\begin{equation}
\tilde{\varrho}_b^A(\tau_A)=\begin{pmatrix}
|C_H|^2 & C_HC_V^*\kappa_+(\tau_A) \\
C_H^*C_V\kappa_+(\tau_A)^* & |C_V|^2
\end{pmatrix},
\label{loc_id_rho_b}
\end{equation}
where
\begin{equation}
\kappa_\pm(\tau_A)=\frac{1\pm e^{-(1-K)\Delta\tau_f^2}\cosh[(1-K)\Delta\tau_f\tau_A]}{1\pm e^{-(1-K)\Delta\tau_f^2}}e^{i\eta\tau_A-\frac{1}{2}\tau_A^2}.
\label{kappa_pm}
\end{equation}
According to the Copenhagen interpretation, the states~\eqref{loc_id_rho_c} and~\eqref{loc_id_rho_b} do not exist until the corresponding photons have already been observed by the detector(s), and so, Eqs.~\eqref{loc_id_rho_c}--\eqref{kappa_pm} actually describe the photons’ average contributions to single-photon polarization tomography when they are projected on Alice and Bob (coincidence), or both on Alice (bunching). Naturally, bunching on Bob's side does not contribute to the state that Alice obtains by averaging over multiple \textit{c} and \textit{b} photons. With an ideal photodetector, that state is the convex combination
\begin{align}
\tilde{\varrho}^A(\tau_A)&=P_c\tilde{\varrho}_c^A(\tau_A)+2P_b^A\tilde{\varrho}_b^A(\tau_A)
\label{prot1_1}\\
&=P_c\tilde{\varrho}_c^A(\tau_A)+P_b\tilde{\varrho}_b^A(\tau_A)
\label{prot1_2}\\
&=\begin{pmatrix}
|C_{H}|^2 & C_HC_V^*\kappa(\tau_A)\\
C_H^*C_V\kappa(\tau_A)^* & |C_V|^2
\end{pmatrix}
\label{prot1_3},
\end{align}
where
\begin{equation}
\kappa(\tau_A)=e^{i\eta\tau_A-\frac{1}{2}\tau_A^2}.
\label{kappa_id}
\end{equation}
There is no information about $K$ or $\Delta\tau_f$ in $\kappa(\tau_A)$ which, quite interestingly, is just the common decoherence function related to Gaussian single-photon dephasing~\cite{sina,partitions,mz_nm}.

\begin{figure}[t!]
\centering
\includegraphics[width=.8\linewidth]{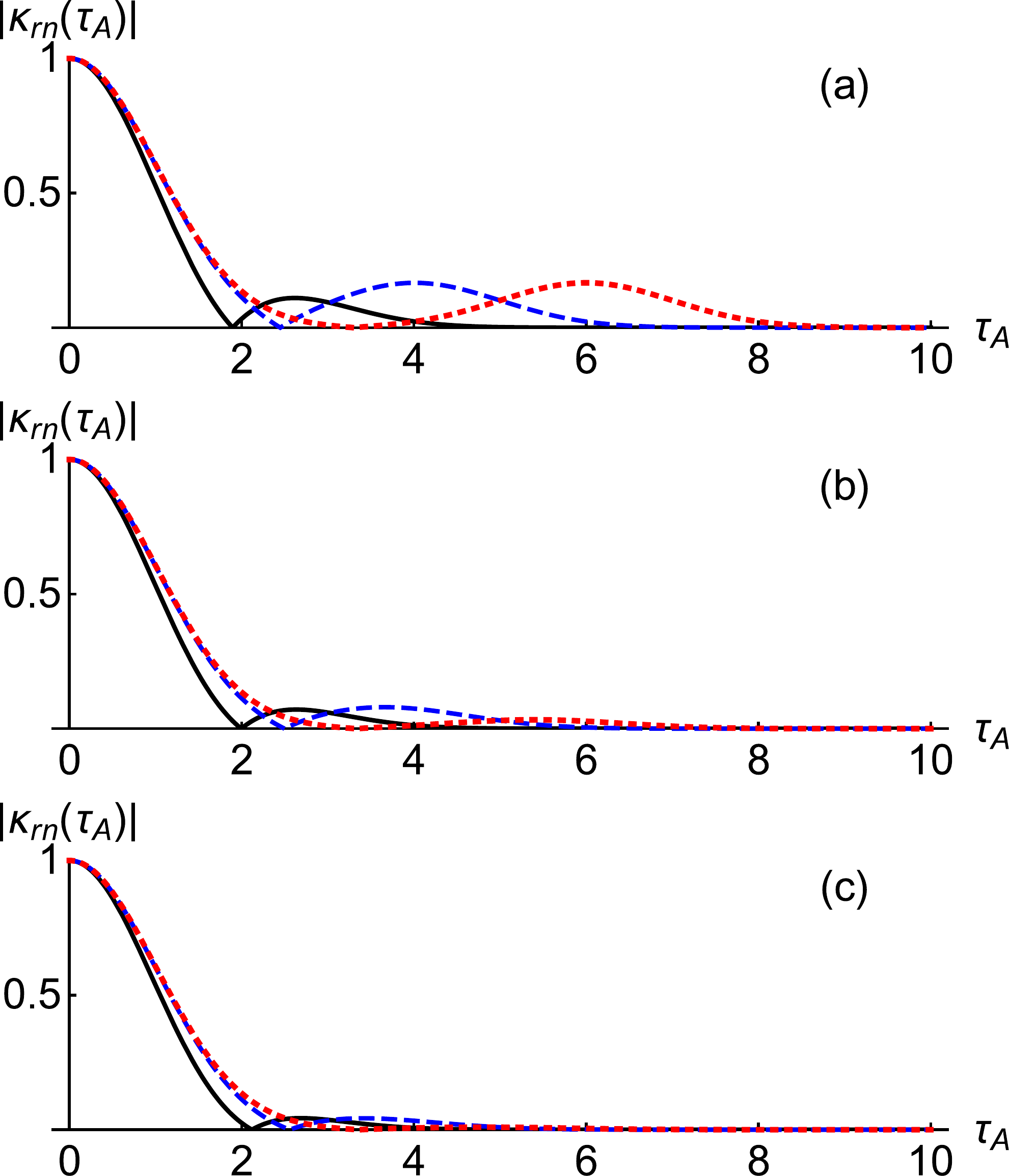}
\caption{(Color online) Absolute values of the renormalized decoherence function $\kappa_{rn}$ as functions of the unitless interaction time $\tau_A$. Black, solid: $|\Delta\tau_f|=1$; Blue, dashed: $|\Delta\tau_f|=2$; Red, dotted: $|\Delta\tau_f|=3$; (a): $K=-1.0$; (b): $K=-0.8$; (c): $K=-0.6$.}
\label{Coherences}
\end{figure}

We now assume that the dead time of Alice's photodetector is great enough to filter out every second \textit{b} photon [cf. Eq.~\eqref{dtc_3} in Appendix C]. Consequently, the renormalized ($rn$) state of Alice's polarization qubit reads
\begin{align}
\tilde{\varrho}_{rn}^A(\tau_A)&=\frac{P_c\tilde{\varrho}_c^A(\tau_A)+P_b^A\tilde{\varrho}_b^A(\tau_A)}{\text{tr}[P_c\tilde{\varrho}_c^A(\tau_A)+P_b^A\tilde{\varrho}_b^A(\tau_A)]}
\label{prot2_1}\\
&=\begin{pmatrix}
|C_H|^2 & C_HC_V^*\kappa_{rn}(\tau_A) \\
C_H^*C_V\kappa_{rn}(\tau_A)^* & |C_V|^2
\end{pmatrix},
\label{prot2_2}
\end{align}
where the decoherence function is
\begin{equation}
\kappa_{rn}(\tau_A)=\frac{3-e^{-(1-K)\Delta\tau_f^2}\cosh[(1-K)\Delta\tau_f\tau_A]}{3-e^{-(1-K)\Delta\tau_f^2}}e^{i\eta\tau_A-\frac{1}{2}\tau_A^2}.
\label{kappa_rn}
\end{equation}
We have plotted the absolute values of $\kappa_{rn}(\tau_A)$ in Fig.~\ref{Coherences} with different values of $K$ and $|\Delta\tau_f|$, which Alice can estimate by fitting $\kappa_{rn}(\tau_A)$ to her measurement data. Together, the height of the ``recoherence peak" (i.e., the peak corresponding to the reviving coherences) gives an estimate for $K$ and its position for $|\Delta\tau_f|$. Clearly, our method works best for the initial polarization states that maximize the local coherences, e.g., $|++\rangle$, and $K\approx-1$. Nevertheless, it is interesting to notice that $\kappa_{rn}(\tau_A)\to\kappa(\tau_A)$ as $K\to1$. This results from the decreasing contribution of the \textit{c} photons and the fact that more and more similar frequency states are being superposed in the beam splitter, which makes $\kappa_+(\tau_A)$ approach $\kappa(\tau_A)$.

We note that the decoherence function $\kappa(\tau_A)$ [Eq.~\eqref{kappa_id}] is also obtained more generally without the simplifying assumption of initially separable and identical polarization states [cf. Eq.~\eqref{general_single_photon_state}]. However, we would not obtain the unique decoherence function $\kappa_{rn}(\tau_A)$ [Eq.~\eqref{kappa_rn}] without the assumption. Moreover, separability rules out the Bell states that would invalidate our method by zero local coherences.

\section{Distinguishing between coincidence and bunching events}

As was observed in the previous section, the polarization states carried by the \textit{c} and \textit{b} photons behave differently in a single-qubit dephasing channel. This suggests that Alice could distinguish between coincidence and bunching events without communicating with Bob---at least up to some probability that we aim to find in this section. We assume that we are well outside the HOM dip, i.e., $(1-K)\Delta\tau_f^2>>0$, meaning that $P_c\approx P_b\approx1/2$. Otherwise the task of distinguishing between coincidence and bunching events would become biased or even meaningless.

In particular, we wish to maximize the trace distance measuring the distinguishability of two states. For the \textit{c} and \textit{b} photons' polarization states, the trace distance is given by
\begin{align}
D_{tr}\big(\tilde{\varrho}_c^A(\tau_A),\tilde{\varrho}_b^A(\tau_A)\big)&=\frac{1}{2}\text{tr}|\tilde{\varrho}_c^A(\tau_A)-\tilde{\varrho}_b^A(\tau_A)|\\
&\nonumber\approx|C_{HH}(C_{HV}^*e^{-\frac{1}{2}\gamma_+}+C_{VH}^*e^{-\frac{1}{2}\gamma_-})\\
&\hspace{5pt}+C_{VV}^*(C_{HV}e^{-\frac{1}{2}\gamma_+}+C_{VH}e^{-\frac{1}{2}\gamma_-})|,
\end{align}
where $\gamma_\pm=\Delta\tau_f^2-2K\Delta\tau_f(\Delta\tau_f\pm\tau_A)+(\Delta\tau_f\pm\tau_A)^2$. In the case of separable and identical initial polarization states, we obtain
\begin{align}
\nonumber D_{tr}\big(\tilde{\varrho}_c^A(\tau_A),\tilde{\varrho}_b^A(\tau_A)\big)=&|C_H||C_V|e^{-(1-K)\Delta\tau_f^2-\frac{1}{2}\tau_A^2}\\
&\times2\cosh\big((1-K)\Delta\tau_f\tau_A\big),
\end{align}
which reaches the maximum value of $1/2$ when $|C_H|=|C_V|=1/\sqrt{2}$, $K=-1$, and $\tau_A=\pm2\Delta\tau_f$. However, we can do better by relaxing the assumption of separability. Noticing that when $e^{-\frac{1}{2}\gamma_+}=1$, $e^{-\frac{1}{2}\gamma_-}\approx0$ (and vice versa), we can focus on the $\gamma_+$ terms and fix $C_{VH}=0$. Simplifying the problem by letting $C_{\lambda\lambda'}\in\mathds{R_+}$, we obtain the trace distance $1/\sqrt{2}$ when $C_{HH}=C_{VV}=1/2$, $C_{HV}=1/\sqrt{2}$, $K=-1$, and $\tau_A=-2\Delta\tau_f$. With these restrictions, but keeping the interaction time $\tau_A$ free, the polarization states corresponding to \textit{c} and \textit{b} photons read, respectively,
\begin{equation}
\tilde{\varrho}_c^A(\tau_A)=\frac{1}{2}\begin{pmatrix}
1 & \nu_-(\tau_A) \\
\nu_-(\tau_A)^* & 1
\end{pmatrix}
\label{max_dtr_c}
\end{equation}
and
\begin{equation}
\tilde{\varrho}_b^A(\tau_A)=\frac{1}{2}\begin{pmatrix}
1 & \nu_+(\tau_A) \\
\nu_+(\tau_A)^* & 1
\end{pmatrix},
\label{max_dtr_b}
\end{equation}
where
\begin{equation}
\nu_\pm(\tau_A)=\frac{1}{\sqrt{2}}e^{i\eta\tau_A}\Big[e^{-\frac{1}{2}\tau_A^2} \pm e^{-\frac{1}{2}(\tau_A+2\Delta\tau_f)^2}\Big].
\label{nu_pm}
\end{equation}

\begin{figure*}[t!]
\begin{minipage}{\textwidth}
\includegraphics[width=.8\textwidth]{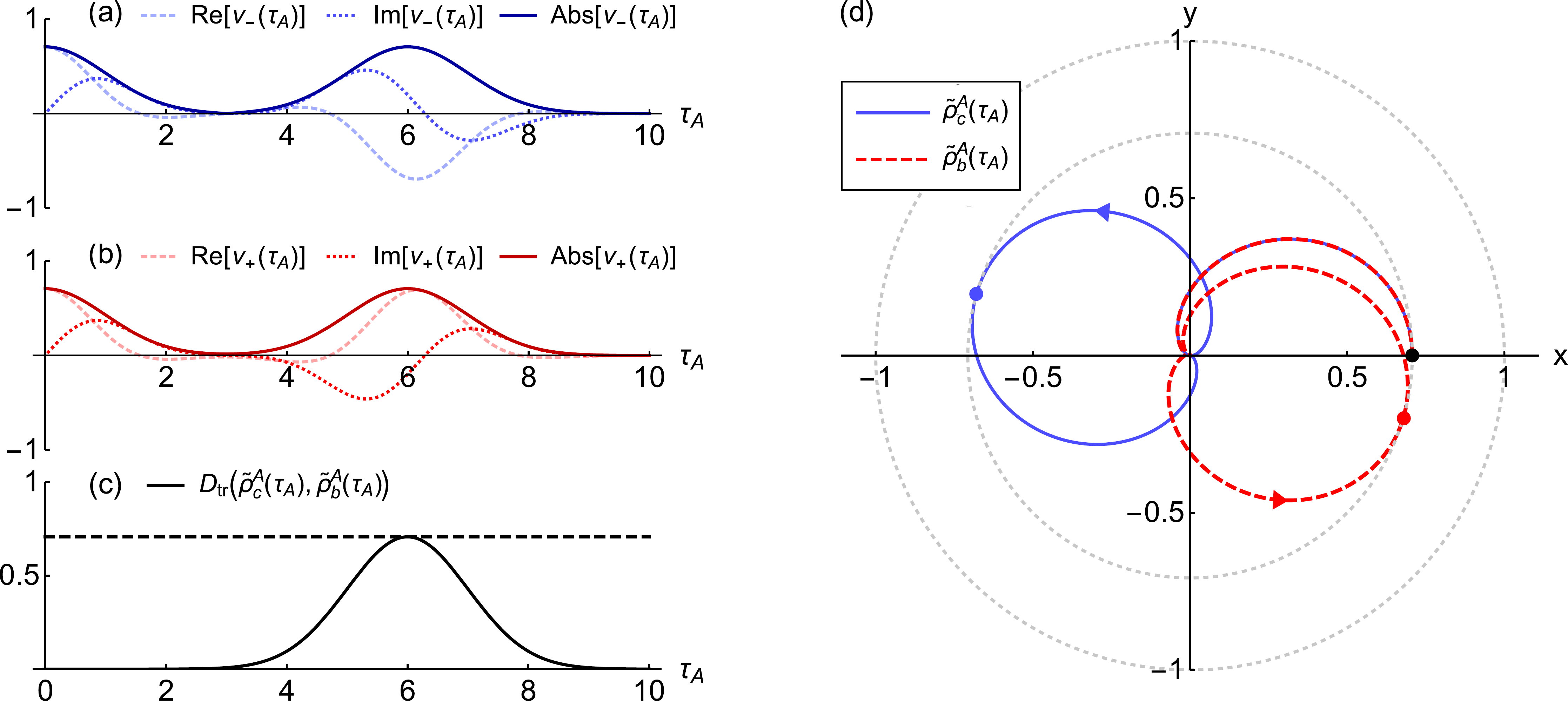}
\caption{(Color online) Dephasing dynamics of the \textit{c} and \textit{b} photons, when the initial polarization state is $(|\Phi^+\rangle+|HV\rangle)/\sqrt{2}$. (a) Real (light-blue, dashed) and imaginary parts (blue, dotted) and the absolute value (dark-blue, solid) of $\nu_-(\tau_A)$. (b) Real (light-red, dashed) and imaginary parts (red, dotted) and the absolute value (dark-red, solid) of $\nu_+(\tau_A)$. (c) Trace distance between $\tilde{\varrho}_c^A(\tau_A)$ and $\tilde{\varrho}_b^A(\tau_A)$. (d) Trajectories of $\tilde{\varrho}_c^A(\tau_A)$ (blue, solid) and $\tilde{\varrho}_b^A(\tau_A)$ (red, dashed) on the xy-plane of the Bloch ball. We have used the parameter values $\Delta\tau_f=-3$ and, for illustrative purposes, $\eta=1$.}
\label{Differences}
\end{minipage}
\end{figure*}

In Fig.~\ref{Differences}, we have plotted $\nu_\pm(\tau_A)$, $D_{tr}\big(\tilde{\varrho}_c^A(\tau_A),\tilde{\varrho}_b^A(\tau_A)\big)$, and the trajectories of the \textit{c} and \textit{b} photons' polarization states inside the Bloch ball, when the initial polarization state is $(|\Phi^+\rangle+|HV\rangle)/\sqrt{2}$. From Fig.~\ref{Differences}, we see that the trajectories of the different polarization states coincide until they first become fully mixed, where they then split up on opposite directions of momentary recoherence. Remarkably, the states achieve the same purity as where they started from. At these local maxima, it is possible for Alice to maximize her probability to correctly identify the \textit{c} and \textit{b} photons without communicating with Bob, i.e., correctly guess if also Bob receives (or received) a photon.

After applying dephasing noise for the duration of $\tau_A=-2\Delta\tau_f$, Alice could, e.g., rotate her states $\pi/2$ around the axis $\hat{n}=(\sin\varphi,\cos\varphi,0)$, where $\varphi=-2\eta\Delta\tau_f$, and use a polarizing beam splitter; first, the rotation operator
\begin{equation}
R_{\hat{n}}(\pi/2)=\frac{1}{\sqrt{2}}\begin{pmatrix}
1 & -e^{i\varphi} \\
e^{-i\varphi} & 1
\end{pmatrix}
\label{rotation}
\end{equation}
transforms the states~\eqref{max_dtr_c} and~\eqref{max_dtr_b} into
\begin{equation}
R_{\hat{n}}(\pi/2)\tilde{\varrho}_c^A(-2\Delta\tau_f)R_{\hat{n}}(\pi/2)^\dagger=\frac{1}{2\sqrt{2}}\begin{pmatrix}
\sqrt{2}+1 & 0 \\
0 & \sqrt{2}-1
\end{pmatrix}
\label{rot_c}
\end{equation}
and
\begin{equation}
R_{\hat{n}}(\pi/2)\tilde{\varrho}_b^A(-2\Delta\tau_f)R_{\hat{n}}(\pi/2)^\dagger=\frac{1}{2\sqrt{2}}\begin{pmatrix}
\sqrt{2}-1 & 0 \\
0 & \sqrt{2}+1
\end{pmatrix}.
\label{rot_b}
\end{equation}
Operating with the rotation matrix $R_{\hat{n}}(\pi/2)$ like this is justified, because the transformations induced by $R_{\hat{n}}(\pi/2)$ on the creation operators commute with the projectors~\eqref{c_proj} and~\eqref{b_proj}.

Then, after separating the $H$ and $V$ components with a polarizing beam splitter, 85.4 $\%$ of the photons in Alice's $H$ branch are \textit{c} photons while 14.6 $\%$ are \textit{b} photons, and vice versa in the $V$ branch, given that $P_c\approx P_b\approx1/2$. Hence, Alice can achieve the success rate of 85.4 $\%$ by guessing that Bob (1) receives a photon if a detector in her $H$ branch clicks, and (2) does \textit{not} receive a photon if a detector in her $V$ branch clicks. It is interesting to notice that even with a separable initial polarization state and $K=0$ it is possible to achieve a success rate higher than 50 $\%$ in a task that is nonlocal by its nature. This is due to the beam splitter, where the photons interact and gain different phase factors with coincidence and bunching.

Going back to the start, our protocol can be directly applied in engineering of the HOM dip, as it constitutes a new kind of delayed choice quantum eraser. Delayed choice quantum erasure refers to erasing the distinguishing information of the input paths \textit{after} the beam splitter and thus reviving the interference pattern. Traditionally, in the case of HOM interference, orthogonal polarizations mark the input paths, while the HOM dip is revived with polarizers~\cite{delayed_choice}. Our results suggest that the same is possible with a distinct path difference marking the input paths. Say, $(1-K)\Delta\tau_f^2>>0$, so that there is no interference at the beam splitter. By scanning across the width of the recoherence peak with different interaction times and comparing her photon counts with Bob's counts, Alice can reconstruct the HOM dip. Of course, Bob can also implement the above-described operations. Different interaction times $\tau_A$ and $\tau_B$ and choices on whether to collect photons from the $H$ or $V$ branch allow Alice and Bob to create multiple distinct HOM dips. However, one must be careful how to define the coincidence probability here. For example, it might happen that the photon pair is projected on Alice and Bob’s $H$ branches, although Bob is collecting his photons from the $V$ branch. As such cases incorrectly contribute to the overall coincidence rate, it would be more appropriate to talk about \textit{pseudo}-HOM dips.

\section{Conclusions}

In this paper, we have expanded the recently introduced concept of open system interferometer~\cite{mz_nm} by considering interacting polarization and frequency degrees of freedom of two photons. To better understand the role of initial frequency correlations, we have used the bivariate Gaussian distribution. In most previous works related to HOM interference, dephasing has been neglected because the birefringence of air and state-of-the-art optical fibres is practically zero. Thus, our setting may appear artificial at first. However, it turns out that amplified noise within the HOM interferometer has a wide range of interesting applications.

We have shown that dephasing provides various ways to engineer the HOM dip. For example, the larger the refractive index, the narrower the HOM dip can be achieved as a function of the interaction time difference. We have also presented an alternative method to construct nonlocal and local Bell states corresponding to coincidence and bunching, respectively, that is based on compensating the path difference before the beam splitter with dephasing after the beam splitter. In a special case, the nonlocal and local Bell states coincide and we obtain the success rate of 1.

In the rest of the paper, we have concentrated on how to approach two nonlocal tasks locally, i.e., without Alice and Bob comparing their photon count statistics with each other. First, we have considered the task of evaluating the correlation coefficient and path difference by single-photon tomography and with the help of sufficiently long dead time of the photodetector. To the best of our knowledge, the dead time has not been discussed before in this particular context. Secondly, we have shown how to improve the probability to distinguish between coincidence and bunching events without an actual coincidence counter. Based on these results, we have proposed a new kind of delayed choice quantum eraser.

The complexity of our model increases rapidly when employing different frequency distributions, wave plates, and more beam splitters. Still, the results arising from our simple setup already highlight the fact that, to better understand biphoton interference, one needs to consider the overall polarization-frequency state. In general, different noise configurations within the HOM interferometer constitute a very promising resource for different quantum information tasks. We hope that our work will also inspire further research on combined continuous- and discrete-time open system dynamics, described here by dephasing and beam splitter, respectively.

\acknowledgements

This work was financially supported by the Magnus Ehrnrooth Foundation and the University of Turku Graduate School (UTUGS).


\onecolumngrid

\subsection*{\label{sec:appendix_a}Appendix A: Calculating the biphoton polarization states corresponding to coincidence and bunching}

\setcounter{equation}{0}
\renewcommand{\theequation}{A\arabic{equation}}

Here, we calculate the biphoton density matrices of polarization corresponding to coincidence and bunching events. For generality, we account for dephasing on paths 0, 1, A, and B, and the difference in the duration of free evolution on paths 0 and 1. In the case of coincidence, Alice and Bob share the non-normalized polarization-frequency state $\Pi_c=\hat{P}_A\otimes\hat{P}_B|\psi_{out}\rangle\langle\psi_{out}|\hat{P}_A\otimes\hat{P}_B$, where the state $|\psi_{out}\rangle$ is given by Eq.~\eqref{output_state}, and the projectors $\hat{P}_A$ and $\hat{P}_B$ are given by Eq.~\eqref{c_proj}. The density matrix elements $\langle\xi\xi'|\varrho_c|\lambda\lambda'\rangle$ of the normalized bipartite polarization state $\varrho_c$ are then given by
\begin{equation}
\langle\xi\xi'|\varrho_c|\lambda\lambda'\rangle=\frac{1}{P_c}\int d\omega_0d\omega_1\langle0_{ab}|\hat{a}_\xi(\omega_0)\hat{b}_{\xi'}(\omega_1)\Pi_c\hat{a}_\lambda^\dagger(\omega_0)\hat{b}_{\lambda'}^\dagger(\omega_1)|0_{ab}\rangle,
\label{coin_elements}
\end{equation}
where $P_c$ is the coincidence probability given by Eq.~\eqref{P_c_3}, and $\xi,\xi',\lambda,\lambda'\in\{H,V\}$. The resulting matrix elements of $\varrho_c$ read
\begin{align}
\langle\lambda\lambda|\varrho_c|\lambda\lambda\rangle&=\frac{|C_{\lambda\lambda}|^2}{2P_c}\big[1-e^{-(1-K)\Delta\tau_{\lambda\lambda}^2}\big],\hspace{5pt}\lambda\in\{H,V\},\\
\nonumber\langle\lambda\lambda'|\varrho_c|\lambda\lambda'\rangle&=\frac{1}{4P_c}\Big\{|C_{HV}|^2+|C_{VH}|^2-2|C_{HV}||C_{VH}|e^{-\frac{1}{2}(\Delta\tau_{HH}^2-2K\Delta\tau_{HH}\Delta\tau_{VV}+\Delta\tau_{VV}^2)}\\
&\hspace{130pt}\times\cos[\eta(\tau_0-\tau_1)+\theta_{HV}-\theta_{VH}]\Big\},\hspace{5pt}\lambda,\lambda'\in\{H,V\},\hspace{5pt}\lambda\neq\lambda',\\
\nonumber\langle HH|\varrho_c|HV\rangle&=\frac{C_{HH}}{4P_c}\Big\{C_{HV}^*e^{i\eta(\tau_1+\tau_B)}\Big(e^{-\frac{1}{2}(\tau_1+\tau_B)^2}-e^{-\frac{1}{2}[\Delta\tau_{HH}^2-2K\Delta\tau_{HH}(\Delta\tau_{HV}+\tau_B)+(\Delta\tau_{HV}+\tau_B)^2]}\Big)\\
&\hspace{33pt}+C_{VH}^*e^{i\eta(\tau_0+\tau_B)}\Big(e^{-\frac{1}{2}(\tau_0+\tau_B)^2}-e^{-\frac{1}{2}[\Delta\tau_{HH}^2-2K\Delta\tau_{HH}(\Delta\tau_{VH}-\tau_B)+(\Delta\tau_{VH}-\tau_B)^2]}\Big)\Big\},\\
\nonumber\langle HH|\varrho_c|VH\rangle&=\frac{C_{HH}}{4P_c}\Big\{C_{HV}^*e^{i\eta(\tau_1+\tau_A)}\Big(e^{-\frac{1}{2}(\tau_1+\tau_A)^2}-e^{-\frac{1}{2}[\Delta\tau_{HH}^2-2K\Delta\tau_{HH}(\Delta\tau_{HV}+\tau_A)+(\Delta\tau_{HV}+\tau_A)^2]}\Big)\\
&\hspace{33pt}+C_{VH}^*e^{i\eta(\tau_0+\tau_A)}\Big(e^{-\frac{1}{2}(\tau_0+\tau_A)^2}-e^{-\frac{1}{2}[\Delta\tau_{HH}^2-2K\Delta\tau_{HH}(\Delta\tau_{VH}-\tau_A)+(\Delta\tau_{VH}-\tau_A)^2]}\Big)\Big\},
\end{align}
\begin{align}
\nonumber\langle HV|\varrho_c|VV\rangle&=\frac{C_{VV}^*}{4P_c}\Big\{C_{HV}e^{i\eta(\tau_0+\tau_A)}\Big(e^{-\frac{1}{2}(\tau_0+\tau_A)^2}-e^{-\frac{1}{2}[\Delta\tau_{VV}^2-2K\Delta\tau_{VV}(\Delta\tau_{HV}+\tau_A)+(\Delta\tau_{HV}+\tau_A)^2]}\Big)\\
&\hspace{30pt}+C_{VH}e^{i\eta(\tau_1+\tau_A)}\Big(e^{-\frac{1}{2}(\tau_1+\tau_A)^2}-e^{-\frac{1}{2}[\Delta\tau_{VV}^2-2K\Delta\tau_{VV}(\Delta\tau_{VH}-\tau_A)+(\Delta\tau_{VH}-\tau_A)^2]}\Big)\Big\},\\
\nonumber\langle VH|\varrho_c|VV\rangle&=\frac{C_{VV}^*}{4P_c}\Big\{C_{HV}e^{i\eta(\tau_0+\tau_B)}\Big(e^{-\frac{1}{2}(\tau_0+\tau_B)^2}-e^{-\frac{1}{2}[\Delta\tau_{VV}^2-2K\Delta\tau_{VV}(\Delta\tau_{HV}+\tau_B)+(\Delta\tau_{HV}+\tau_B)^2]}\Big)\\
&\hspace{30pt}+C_{VH}e^{i\eta(\tau_1+\tau_B)}\Big(e^{-\frac{1}{2}(\tau_1+\tau_B)^2}-e^{-\frac{1}{2}[\Delta\tau_{VV}^2-2K\Delta\tau_{VV}(\Delta\tau_{VH}-\tau_B)+(\Delta\tau_{VH}-\tau_B)^2]}\Big)\Big\},\\
\nonumber\langle HH|\varrho_c|VV\rangle&=\frac{C_{HH}C_{VV}^*}{4P_c}e^{i\eta(\tau_0+\tau_1+\tau_A+\tau_B)}\Big\{e^{-\frac{1}{2}[(\tau_0+\tau_A)^2+2K(\tau_0+\tau_A)(\tau_1+\tau_B)+(\tau_1+\tau_B)^2]}\\
\nonumber&\hspace{121pt}+e^{-\frac{1}{2}[(\tau_0+\tau_B)^2+2K(\tau_0+\tau_B)(\tau_1+\tau_A)+(\tau_1+\tau_A)^2]}\\
\nonumber&\hspace{121pt}-e^{-\frac{1}{2}[(\Delta\tau_{HV}+\tau_A)^2-2K(\Delta\tau_{HV}+\tau_A)(\Delta\tau_{VH}-\tau_B)+(\Delta\tau_{VH}-\tau_B)^2]}\\
&\hspace{121pt}-e^{-\frac{1}{2}[(\Delta\tau_{HV}+\tau_B)^2-2K(\Delta\tau_{HV}+\tau_B)(\Delta\tau_{VH}-\tau_A)+(\Delta\tau_{VH}-\tau_A)^2]}\Big\},\\
\nonumber\langle HV|\varrho_c|VH\rangle&=\frac{1}{4P_c}\Big\{C_{HV}C_{VH}^*e^{i\eta(\tau_0-\tau_1+\tau_A-\tau_B)-\frac{1}{2}[(\tau_0+\tau_A)^2-2K(\tau_0+\tau_A)(\tau_1+\tau_B)+(\tau_1+\tau_B)^2]}\\
\nonumber&\hspace{27pt}+C_{HV}^*C_{VH}e^{i\eta(-\tau_0+\tau_1+\tau_A-\tau_B)-\frac{1}{2}[(\tau_0+\tau_B)^2-2K(\tau_0+\tau_B)(\tau_1+\tau_A)+(\tau_1+\tau_A)^2]}\\
\nonumber&\hspace{27pt}-|C_{HV}|^2e^{i\eta(\tau_A-\tau_B)-\frac{1}{2}[(\Delta\tau_{HV}+\tau_A)^2-2K(\Delta\tau_{HV}+\tau_A)(\Delta\tau_{HV}+\tau_B)+(\Delta\tau_{HV}+\tau_B)^2]}\\
&\hspace{27pt}-|C_{VH}|^2e^{i\eta(\tau_A-\tau_B)-\frac{1}{2}[(\Delta\tau_{VH}-\tau_A)^2-2K(\Delta\tau_{VH}-\tau_A)(\Delta\tau_{VH}-\tau_B)+(\Delta\tau_{VH}-\tau_B)^2]}\Big\},
\end{align}
where the path difference has been included in the shorthand notations $\Delta\tau_{\lambda\lambda'}=\sigma(t_{0f}+n_{0\lambda}t_0-t_{1f}-n_{1\lambda'}t_1)$, $\lambda,\lambda'\in\{H,V\}$, unlike in the main text.

In the case of bunching on Alice's side, the non-normalized polarization-frequency state becomes $\Pi_b^A=\hat{P}_{AA}|\psi_{out}\rangle\langle\psi_{out}|\hat{P}_{AA}$, where $\hat{P}_{AA}$ is given by Eq.~\eqref{b_proj}. The density matrix elements $\langle\xi\xi'|\varrho_b^A|\lambda\lambda'\rangle$ of the normalized polarization state $\varrho_b^A$ are then given by
\begin{equation}
\langle\xi\xi'|\varrho_b^A|\lambda\lambda'\rangle=\frac{1}{2P_b^A}\int d\omega_0d\omega_1\langle0_{ab}|\hat{a}_\xi(\omega_0)\hat{a}_{\xi'}(\omega_1)\Pi_b^A\hat{a}_\lambda^\dagger(\omega_0)\hat{a}_{\lambda'}^\dagger(\omega_1)|0_{ab}\rangle,
\label{bunch_elements}
\end{equation}
where the factor of $1/2$ is a normalization constant coming from the commutation relation of bosonic creation and annihilation operators, and $P_b^A=(1-P_c)/2$. The density matrix elements of $\varrho_b^A$ now read
\begin{align}
\langle\lambda\lambda|\varrho_b^A|\lambda\lambda\rangle&=\frac{|C_{\lambda\lambda}|^2}{4P_b^A}\big[1+e^{-(1-K)\Delta\tau_{\lambda\lambda}^2}\big],\hspace{5pt}\lambda\in\{H,V\},\\
\nonumber\langle\lambda\lambda'|\varrho_b^A|\lambda\lambda'\rangle&=\frac{1}{8P_b^A}\Big\{|C_{HV}|^2+|C_{VH}|^2+2|C_{HV}||C_{VH}|e^{-\frac{1}{2}(\Delta\tau_{HH}^2-2K\Delta\tau_{HH}\Delta\tau_{VV}+\Delta\tau_{VV}^2)}\\
&\hspace{130pt}\times\cos[\eta(\tau_0-\tau_1)+\theta_{HV}-\theta_{VH}]\Big\},\hspace{5pt}\lambda,\lambda'\in\{H,V\},\hspace{5pt}\lambda\neq\lambda',\\
\nonumber\langle HH|\varrho_b^A|\lambda\lambda'\rangle&=\frac{C_{HH}}{8P_b^A}\Big\{C_{HV}^*e^{i\eta(\tau_1+\tau_A)}\Big(e^{-\frac{1}{2}(\tau_1+\tau_A)^2}+e^{-\frac{1}{2}[\Delta\tau_{HH}^2-2K\Delta\tau_{HH}(\Delta\tau_{HV}+\tau_A)+(\Delta\tau_{HV}+\tau_A)^2]}\Big)\\
&\hspace{33pt}+C_{VH}^*e^{i\eta(\tau_0+\tau_A)}\Big(e^{-\frac{1}{2}(\tau_0+\tau_A)^2}+e^{-\frac{1}{2}[\Delta\tau_{HH}^2-2K\Delta\tau_{HH}(\Delta\tau_{VH}-\tau_A)+(\Delta\tau_{VH}-\tau_A)^2]}\Big)\Big\},\hspace{5pt}\lambda,\lambda'\in\{H,V\},\hspace{5pt}\lambda\neq\lambda',\\
\nonumber\langle \lambda\lambda'|\varrho_b^A|VV\rangle&=\frac{C_{VV}^*}{8P_b^A}\Big\{C_{HV}e^{i\eta(\tau_0+\tau_A)}\Big(e^{-\frac{1}{2}(\tau_0+\tau_A)^2}+e^{-\frac{1}{2}[\Delta\tau_{VV}^2-2K\Delta\tau_{VV}(\Delta\tau_{HV}+\tau_A)+(\Delta\tau_{HV}+\tau_A)^2]}\Big)\\
&\hspace{30pt}+C_{VH}e^{i\eta(\tau_1+\tau_A)}\Big(e^{-\frac{1}{2}(\tau_1+\tau_A)^2}+e^{-\frac{1}{2}[\Delta\tau_{VV}^2-2K\Delta\tau_{VV}(\Delta\tau_{VH}-\tau_A)+(\Delta\tau_{VH}-\tau_A)^2]}\Big)\Big\},\hspace{5pt}\lambda,\lambda'\in\{H,V\},\hspace{5pt}\lambda\neq\lambda',\\
\nonumber\langle HH|\varrho_b^A|VV\rangle&=\frac{C_{HH}C_{VV}^*}{4P_b^A}e^{i\eta(\tau_0+\tau_1+2\tau_A)}\Big\{e^{-\frac{1}{2}[(\tau_0+\tau_A)^2+2K(\tau_0+\tau_A)(\tau_1+\tau_A)+(\tau_1+\tau_A)^2]}\\
&\hspace{112pt}+e^{-\frac{1}{2}[(\Delta\tau_{HV}+\tau_A)^2-2K(\Delta\tau_{HV}+\tau_A)(\Delta\tau_{VH}-\tau_A)+(\Delta\tau_{VH}-\tau_A)^2]}\Big\},
\end{align}
\begin{align}
\nonumber\langle HV|\varrho_b^A|VH\rangle&=\frac{1}{8P_b^A}\Big\{|C_{HV}|^2e^{-(1-K)(\Delta\tau_{HV}+\tau_A)^2}+|C_{VH}|^2e^{-(1-K)(\Delta\tau_{VH}-\tau_A)^2}\\
&\hspace{22pt}+2|C_{HV}||C_{VH}|e^{-\frac{1}{2}[(\tau_0+\tau_A)^2-2K(\tau_0+\tau_A)(\tau_1+\tau_A)+(\tau_1+\tau_A)^2]}\cos[\eta(\tau_0-\tau_1)+\theta_{HV}-\theta_{VH}]\Big\}.
\end{align}
The matrix elements of $\varrho_b^B$ (bunching on Bob's side) are the same, except with the labels B instead of A.

\subsection*{\label{sec:appendix_b}Appendix B: Calculating the single-photon polarization states corresponding to coincidence and bunching}

\setcounter{equation}{0}
\renewcommand{\theequation}{B\arabic{equation}}

Here, we calculate Alice's single-photon polarization states corresponding to coincidence and bunching events. Bob's states are calculated similarly. Once the two-photon state $\varrho_c$ corresponding to coincidence is known, Alice's state $\tilde{\varrho}_c^A$ is obtained by taking partial trace over Bob's photon
\begin{equation}
\tilde{\varrho}_c^A=
\begin{pmatrix}
\langle HH|\varrho_c|HH\rangle + \langle HV|\varrho_c|HV\rangle &
\langle HH|\varrho_c|VH\rangle + \langle HV|\varrho_c|VV\rangle \\
\langle VH|\varrho_c|HH\rangle + \langle VV|\varrho_c|HV\rangle &
\langle VH|\varrho_c|VH\rangle + \langle VV|\varrho_c|VV\rangle
\end{pmatrix}.
\label{rho_c_A}
\end{equation}

In the case of bunching, the single-photon states are obtained by taking partial trace of $\varrho_b^A$ over the other photon. The resulting states are found to be equal
\begin{align}
\tilde{\varrho}_b^A&=
\begin{pmatrix}
\langle HH|\varrho_b^A|HH\rangle + \langle HV|\varrho_b^A|HV\rangle &
\langle HH|\varrho_b^A|VH\rangle + \langle HV|\varrho_b^A|VV\rangle \\
\langle VH|\varrho_b^A|HH\rangle + \langle VV|\varrho_b^A|HV\rangle &
\langle VH|\varrho_b^A|VH\rangle + \langle VV|\varrho_b^A|VV\rangle
\end{pmatrix}\\
&=
\begin{pmatrix}
\langle HH|\varrho_b^A|HH\rangle + \langle VH|\varrho_b^A|VH\rangle &
\langle HH|\varrho_b^A|HV\rangle + \langle VH|\varrho_b^A|VV\rangle \\
\langle HV|\varrho_b^A|HH\rangle + \langle VV|\varrho_b^A|VH\rangle &
\langle HV|\varrho_b^A|HV\rangle + \langle VV|\varrho_b^A|VV\rangle
\end{pmatrix}.
\label{rho_b_A}
\end{align}

If Alice performs single-photon tomography when there is no dephasing prior to the beam splitter, the state that Alice can construct with an ideal photodetector is
\begin{align}
\tilde{\varrho}^A(\tau_A)&=P_c\tilde{\varrho}_c^A(\tau_A)+2P_b^A\tilde{\varrho}_b^A(\tau_A)\\
&=\frac{1}{2}\begin{pmatrix}
1+|C_{HH}|^2-|C_{VV}|^2 & [C_{HH}(C_{HV}^*+C_{VH}^*)+(C_{HV}+C_{VH})C_{VV}^*]\kappa(\tau_A) \\
[C_{HH}^*(C_{HV}+C_{VH})+(C_{HV}^*+C_{VH}^*)C_{VV}]\kappa(\tau_A)^* & 1-|C_{HH}|^2+|C_{VV}|^2
\end{pmatrix},
\label{general_single_photon_state}
\end{align}
where no assumptions about the initial polarization state have been made, and the decoherence function $\kappa(\tau_A)$ is given by Eq.~\eqref{kappa_id}. Interestingly, $\tilde{\varrho}^A(0)$ is just the average of the initial path-wise states, $\tilde{\varrho}^A(0)=(\tilde{\varrho}_0+\tilde{\varrho}_1)/2$.

\subsection*{\label{sec:appendix_c}Appendix C: Deriving the condition for sufficiently long dead time}

\setcounter{equation}{0}
\renewcommand{\theequation}{C\arabic{equation}}

\begin{figure}[t!]
\centering
\includegraphics[width=.6\linewidth]{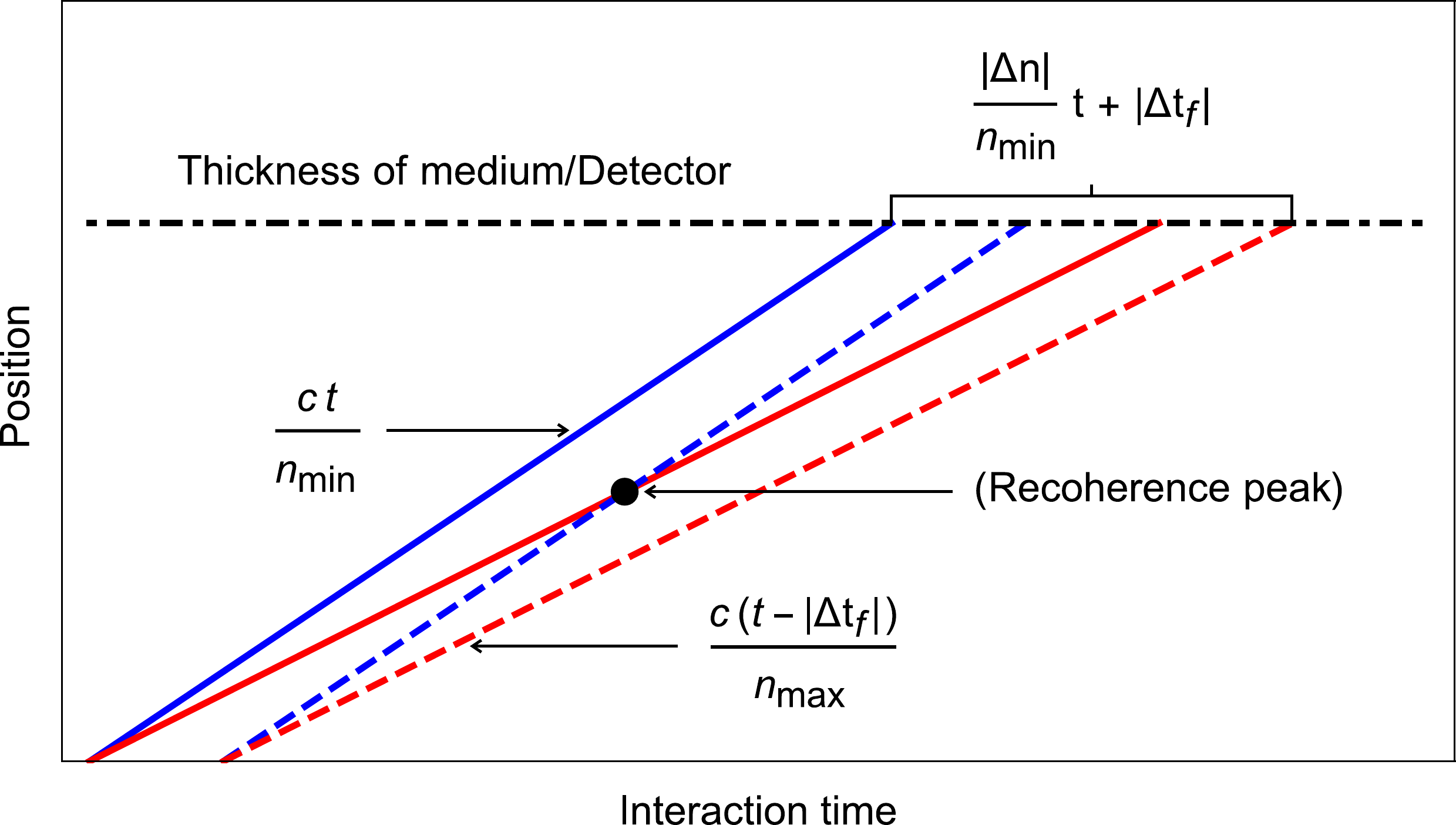}
\caption{(Color online) The fast and slow components of two consecutive \textit{b} photons. The birefringent medium starts from the bottom of the picture and ends at the black dash-dotted line which also represents the photodetector; possible free evolution between the medium and detector need not be taken into account, because that would mean all of the components traveling parallel to each other. Recoherence reaches its maximum value at the intersection of the first slower component (red, solid) and the second faster component (blue, dashed).}
\label{DTC}
\end{figure}

Here, we derive the condition for the dead time of a photodetector to filter out every second \textit{b} photon. Essentially, the detector needs to be off for the time span upon which it could detect the second photon. We have plotted the fast and slow components of two consecutive \textit{b} photons in Fig.~\ref{DTC}. The time span in question is given by $t'-t$, where $t$ ($t'$) is the earliest (latest) point of interaction time where the detector is reached. Denoting the thickness of the medium by $d$, we obtain
\begin{align}
d&=\frac{ct}{\min\{n_H,n_V\}}
\label{dtc_1}\\
&=\frac{c(t'-|\Delta t_f|)}{\max\{n_H,n_V\}}
\label{dtc_2}\\
\Leftrightarrow t'-t&=\frac{|\Delta n|}{\min\{n_H,n_V\}}t+|\Delta t_f|.
\label{dtc_3}
\end{align}
Evidently, the longer the interaction time $t$ is being implemented, the longer the time span there is between the furthest components. Furthermore, for the single-photon tomography to work as proposed in Sec. V, Alice should detect one \textit{c} photon to every \textit{b} photon on average. This means that the \textit{c} photons must not be filtered, which again means that the time between each photon-pair generation must be larger than $|\Delta t_f|$.

Above, we treated the photons as temporally localized particles, although fully correlated or anticorrelated photons are \textit{delocalized} in time. However, our analysis is justified by the fact that detecting one of the photons localizes the other one. In the special case of the real-valued probability amplitude $g(\omega_0,\omega_1)=|g(\omega_0,\omega_1)|$, the (non-normalized) biphoton probability amplitude at time $(s_0,s_1)$ is obtained as the Fourier transform of $g(\omega_0,\omega_1)$~\cite{duality}
\begin{align}
\hat{g}(s_0,s_1)&=\int d\omega_0d\omega_1g(\omega_0,\omega_1)e^{is_0\omega_0}e^{is_1\omega_1}\\
&=\sqrt{8\pi\sigma^2\sqrt{1-K^2}}e^{-\sigma^2(s_0^2+2Ks_0s_1+s_1^2)+i\mu(s_0+s_1)}.
\end{align}
The corresponding (normalized) probability distribution is
\begin{equation}
|\hat{g}(s_0,s_1)|^2=\frac{2\sigma^2\sqrt{1-K^2}}{\pi}e^{-2\sigma^2(s_0^2+2Ks_0s_1+s_1^2)},
\label{temporal_prob_dist}
\end{equation}
and its margins are given by
\begin{equation}
|\hat{g}(s_j)|^2=\sqrt{\frac{2\sigma^2(1-K^2)}{\pi}}e^{-2\sigma^2(1-K^2)s_j^2},
\label{temporal_prob_margin}
\end{equation}
where $j\in\{0,1\}$. Now, the conditional probability distribution of the photon 0 given that the photon 1 has been detected at some time $S_1$ is
\begin{align}
|\hat{g}(s_0|s_1=S_1)|^2&=\frac{|\hat{g}(s_0,S_1)|^2}{|\hat{g}(S_1)|^2}\\
&=\sqrt{\frac{2\sigma^2}{\pi}}e^{-2\sigma^2(s_0+KS_1)^2},
\end{align}
which is clearly localized in time. A more detailed discussion on the localization of frequency-anticorrelated biphotons can be found, e.g., in~\cite{localization}.

\twocolumngrid

\end{document}